**Efficient coding under constraint drives neural systems towards criticality and sloppiness**

He Xiao[1,2], Xinyue Zhao[1,2], Haijun Zhou[1,2], Weikang Wang[1,2*]

[1]Institute of Theoretical Physics, Chinese Academy of Sciences, Beijing 100190, China
[2] School of Physical Sciences, University of Chinese Academy of Sciences, Beijing 100049, China

Correspondence: wangwk@itp.ac.cn, zhouhj@itp.ac.cn

**Abstract**

It is widely accepted that the brain operates near a critical state, characterized by neural avalanches that follow power-law distributions. However, the functional rationale for why neural systems attain criticality remains unclear. Here, we present a theoretical framework that links efficient coding to criticality in neural populations. Using a Gaussian population coding model, we demonstrate that maximizing Fisher information under resource constraints naturally leads to the emergence of soft modes and diverging correlation lengths, which are hallmarks of criticality. By introducing spatial structure, we unify two distinct perspectives of criticality: statistical criticality with diverging correlation lengths and dynamical criticality with critical slowing down as well as bifurcation. Furthermore, this framework provides a natural explanation for the sloppiness observed in neural systems. Numerical simulations confirm that optimization results in power-law response, providing a mechanistic link between efficient coding, sloppiness and the critical brain hypothesis.

## I. INTRODUCTION

The hypothesis that the brain operates near a critical point has gained substantial experimental support over the past two decades. And this criticality can be attributed as self-organized criticality (SOC), emerged from simple models like the sandpile, where a driven system naturally tunes itself to criticality without parameter fine-tuning [1]. And power-law distributions of nerual avalanche across multiple scales is the most important phenonmonon indicating criticality [2-4]. This critical state is theoretically advantageous, offering optimal dynamic range, information transmission, and computational capacity [5-7]. However, besides the advantages, why neural systems have to achieve and maintain criticality remain incompletely understood.

In statistical physics, criticality means diverging correlation length and power law distribution. In dynamical systems, it refers to marginal stability, bifurcation and critical slowing down. These are typically studied separately in the critical brain literature [2]. Unifying them is therefore crucial for elucidating neural dynamics. In this work, we present a theoretical framework for investigating the relationship between these two forms of criticality in neural systems from the unifying perspective of Fisher information.

Fisher information quantifies the sensitivity of a system's internal representation to changes in external stimulus [8]. Increasing Fisher information enhances the system's ability to discriminate stimuli, improving fitness in variable environments. Neural systems must efficiently encode environmental information. Efficient coding theory posits that neural systems maximize mutual information between inputs and representations [9,10]. Fisher information provides a way to estimate the mutual information between a stimulus $s$ and the evoked neural response $x$ [8]:

$$I(s,x) \approx \frac{1}{2}\int p(s) \ln J(s)\, ds$$

where $p(s)$ is the probability distribution of the stimulus and $J(s)$ is the Fisher information. Hence neural system tends maximize Fisher information. Simliarly, living systems that have to increase their fitness to enviroment also maximizes Fisher information due natural selection [11,12]. Fisher information is also closely related to the concept of sloppiness, which refers to the anisotropic sensitivity of parameter space in neural systems[13,14]. In addition, studies have revealed the utility of Fisher information for characterizing criticality and phase transitions [15,16].

By analyzing neural population coding from the perspective of Fisher information, we identify a connection among efficient coding, sloppiness, and criticality. Moreover, both forms of criticality can be unified within this minimal model.

## II. THEORETICAL FRAMEWORK

### A. Fisher Information of Neural Population Coding Model

Consider a neural population of $N$ neurons responding to a continuous stimulus $s \in \mathbb{R}$. Because of trial-to-trial variability, neural population coding is probabilistic. The variability of neural responses can be decomposed into two types of correlations: signal correlations and noise correlations [17].

The neural response vector $x \in \mathbb{R}^N$ is give by

$$x = f(s) + n$$

where $f_i(s)$ is the averaged repsonse and $n_i$ reprensents the response variability of neuron $i$. For simplicity, we consider the case where the stimulus is fixed, so that only noise correlations are present. Noise correlations are affected by synaptic connectivity and shared inputs [18-20].

Therefore, we employ a recurrent network structure to modulate the noise correlations (details are provided below). Studies have shown that learning-related changes in tuning properties and noise correlations can impact discrimination ability and information capacity, with modulation of noise correlations playing a particularly more crucial role [21,22].
We further assume that the conditional neural response follows a Gaussian distribution with covariance matrix $\Sigma$ which is independent of stimulus $s$ [23]. The conditional distribution can then be written as:

$$P(x \mid s) = \sqrt{\frac{1}{(2\pi)^N}} \sqrt{\det(A)} \exp\left(-\frac{1}{2}(x - f(s))^T A (x - f(s))\right)$$

where $A = \Sigma^{-1}$ is the precision matrix. This Gaussian assumption allows analytical treatment while capturing essential features of population coding.

The Fisher information with respect to stimulus $s$ is:

$$J(s) = \left(\frac{df}{ds}\right)^T A \left(\frac{df}{ds}\right)$$

We assume covariance $C$ (hence precision $A$) is independent of $s$ for simplicity. The precision matrix $A$ is positive semidefinite and can be decomposed as:

$$A = U \Lambda U^T, \Lambda = \mathrm{diag}(\lambda_1, \lambda_2, \dots, \lambda_N)$$

where $\lambda_i > 0$ are eigenvalues and $U$ contains eigenvectors $u_i$. Then:

$$J(s) = \sum_{i=1}^{N} \lambda_i v_i^2, v_i = u_i^T g(s), g(s) = \frac{df}{ds}$$

The terms $v_i$ represent projections of the sensitivity vector $g(s)$ onto each eigenvector. In our simulation, the explicit form of $f(s)$ is not necessary;only $g(s)$ is required. Since we use a fixed stimulus, we assume that $g(s)$ is static, i.e., it does not vary with the stimulus.

**B. Maximizing Fisher information under constraints**

In biological systems, maintaining precision requires metabolic energy. We assume energy cost is proportional to total precision:

$$\text{Cost} \propto \sum_i \frac{1}{\sigma_i^2} = \mathrm{Tr}(A)$$

where $\sigma_i^2$ are variances along eigen directions [24]. This trace constraint reflects the metabolic expense of reducing neural variability. Under this constraint, maximizing Fisher information forces the system to allocate most resources to directions with large $v_i^2$, minimizing eigenvalues in other directions.

Fisher information is maximized when the eigenvector corresponding to the largest eigenvalue aligns with $g(s)$, and $\lambda_{\max}$ is as large as possible (Fig. 1a). Without other constraints, $A$ would approach a rank-1 matrix $g(s)g(s)^T$, concentrating all precision in the stimulus-relevant direction (Fig. 1b).

In practice, we add a penalty term $\ln(\det A)$, which is proportional to the negative differential entropy of the Gaussian distribution $P(\mathrm{x}|s)$. It prevents the system from collapsing into a zero-entropy state.

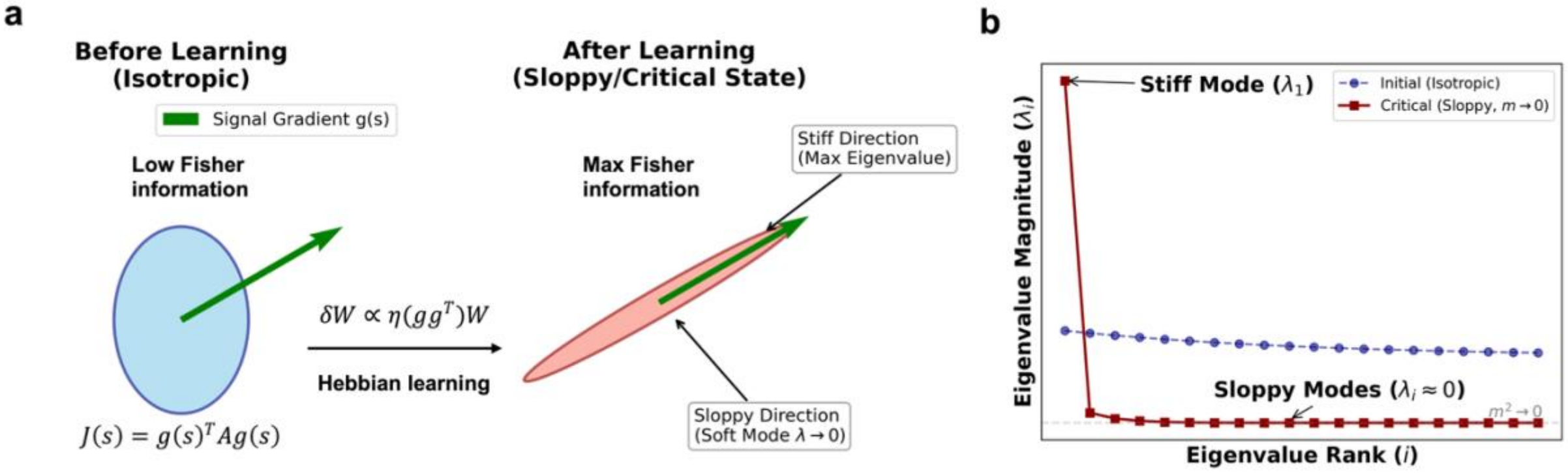

**Figure 1** Variation of precision matrix in learning. (a) The eigenvector corresponding to the largest eigenvalue is aligned with the stimulus sensitivity vector and this largest eigenvalue is growing close to trace of A. Other eigenvalues are shrinked to nearly 0. (b) The eigenvalue spectrums before and after learning.

### C. Learning Rule from Fisher Information Gradient

The gradient of Fisher information suggests a learning rule for the precision matrix $A$. To ensure $A$ remains positive semi-definite during optimization, we employ the decomposition $A = WW^T$. The Fisher Information then becomes $J(s) = g^T W W^T g = ||W^T g||^2$. We seek to maximize $J(s)$ subject to a resource constraint on the total precision $\mathrm{Tr}(A) = ||W||_F^2$ and entropy penalty of $\ln(\det A)$.

The Lagrangian is $\mathcal{L} = g^T W W^T g + \alpha ln(det(A)) - \beta ||W||_F^2$ and the gradient with respect to $W$ is $\frac{\partial \mathcal{L}}{\partial W} = 2(gg^T)W + 2\alpha A^{-1} W - 2\beta W$. In a gradient ascent scheme, the update rule for $W$ emerges as $\Delta W \propto \eta(gg^T W + \alpha A^{-1} W - \beta W)$. Ignoring the terms from constrints for the local learning dynamics, we arrive at the Hebb-like rule:

$$\delta W \propto \eta(gg^T)W$$

This rule is Hebb-like because the term $(gg^T)W$ describes an association between the stimulus sensitivity $g$ and the existing weight structure $W$. Hence the precision matrix and noise correlation can be adjusted with Hebbian learning [25,26].
This formulation maps directly onto the predictive coding architecture[27-29]. In predictive coding, the brain maintains an internal prediction $f(s)$ and processes the residual error or deviation $z = x - f(s)$. If we interpret $W^T$as a transformation matrix, then $\epsilon = W^T z$ represents the whitened prediction error. The decomposition $A = WW^T$ suggests a two-stage process: residue errors are firstly projected by $W^T$, followed by a lateral modulation by $W$ . The two-stage process can be interpreted as the neural network serving two roles: state neurons and error neurons (Fig.2). The learning rule $\delta W \propto gg^T W$ implies that the weights $W$ adapt to align the system's stiffest response axis with the most informative stimumlus dimensions.

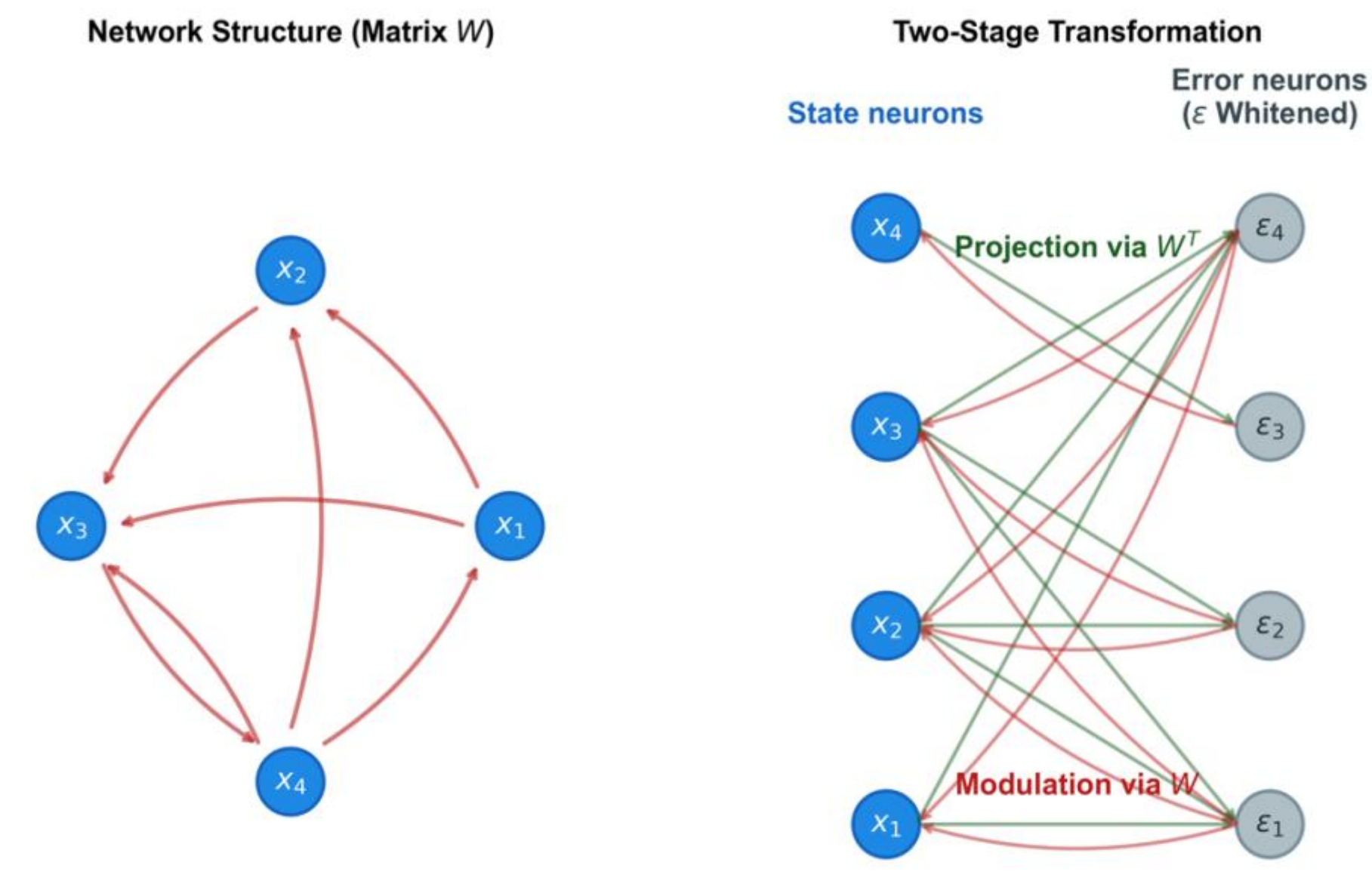

Figure 2 Interpretation the decomposion of precision matrix with predictive coding network architecture. Left: recurrent network structure. Right: two-stage transformation.

## D. Soft Mode and Correlation Length

Neural populations are typically embedded in spatial structures, such as cortical sheets, where structural connectivity are governed by distance-dependent interactions. Here we employ a recurrent network stucture with lateral interactions between neurons defined by a spatial mask $M$ which impose spatial constraints on the structure connectivity $W$. A connection $W_{ij}$ exists only if $M_{ij} = 1$.

For complex networks, the correlation length can be determined from the spectral gap [30]. The smallest eigenvalue $\lambda_{\min}$ of $A$ corresponds to the longest-wavelength mode—the softest mode in the system. In a continuum approximation, the correlation length scales as

$$\xi \propto \frac{1}{\sqrt{\lambda_{\min}}}.$$

However, in heterogeneous networks, the softest mode may become localized. The localization property of an eigenvector $u_{\lambda_i}$ can be quantified by the inverse participation ratio (IPR):

$$\mathrm{IPR}(\lambda_i) = \sum_j u_{\lambda_i,j}^4 \,.$$

If the soft mode is delocalized, each component scales as $O(1/\sqrt{N})$ and IPR $= O(1/N)$ [31-33]. If the mode is localized, its components are concentrated on a few nodes and the IPR remains finite. In the thermodynamic limit $N \to \infty$, a delocalized mode satisfies IPR $\propto N^{-1} \to 0$. Specifically, we also considered a homogeneous network where neurons are arranged on a regular lattice $M$ with local coupling and the precision matrix $A$ becomes banded and approximately translation invariant. In the continuum limit, any such local symmetric operator can be approximated by a power series of the Laplacian $\nabla^2$. If the stimulus senstivity $g(s)$ has spatial smoothness, the learning process effectively smooths the precision matrix over the mask's support. Hence, we model this by treating $A$ as a Gaussian random field (GRF) operator [34,35]. In continuum approximation, the precision matrix becomes a differential operator:

$$A = m^2 - \kappa\nabla^2$$

where $m$ and $\kappa$ are parameters determining the baseline precision and coupling strength. In Fourier space:

$$A(k) = m^2 + \kappa k^2$$

where $m$ acts as a mass term and $\kappa$ controls spatial correlations. The correlation length is:

$$\xi = \sqrt{\kappa}/m$$

The eigenvalues in Fourier modes are $\lambda_k = m^2 + \kappa k^2$, with the smallest eigenvalue $\lambda_{k=0} = m^2$ corresponding to the longest wavelength mode—the softest mode in the system.

### E. Relaxation Dynamics

The relaxation dynamics of the neural system around the fixed point can be approximated by gradient of the log-probability:

$$\frac{dx}{dt} = \nabla_x \log P(x \mid s) = -A\big(x - f(s)\big) = -WW^T\big(x - f(s)\big)$$

The predictive coding arachiture is clear from the dynamics equation. The error neurons compute the weighted projection of the prediction error

$$\epsilon = W^T\big(\mathrm{x} - \mathrm{f}(s)\big)$$

And the state neurons receive feedback from the error neurons

$$\frac{d\mathrm{x}}{dt} = -W\epsilon$$

Defining deviation $z = x - f(s)$:

$$\dot{z} = -Az$$

Discretizing with time step $dt$:

$$z_{t+1} = z_t - dtAz_t = (I - dtA)z_t$$

Define transfter matrix $T = I - dtA$, with eigenvalues $\mu_i = 1 - dt\lambda_i$. The dynamics of $z$ around the fixed point are governed by $T$.

**III. NUMERICAL RESULTS**

We initialize the mask $M$ as a random or small-world graph to model heterogeneous connectivity [36,37]. In this case, the soft mode may become localized due to topological disorder. To prevent localization and promote a global soft mode, we incorporate two additional mechanisms during learning.

1. Dropout. At each learning step, a random subset of connections is dropped i.e., not updated. This stochasticity helps break local traps and encourages diverse connectvity path.

2. Graph Laplacian regularization. We add a regularization term that penalizes weight differences between adjacent neurons:
$$\mathcal{R} = \sum_{p,q} M_{pq} \parallel r_p - r_q \parallel^2,$$
where $r_p$ denotes the $p$-th row of the whitening matrix $W$. This term can be rewritten as $\mathcal{R} = \gamma\, \mathrm{Tr}(W^T L_s W)$, with $\gamma$ the regularization strength and $L_s$ the graph Laplacian of the network (defined from $M$). The corresponding contribution to the gradient of $W$ is $-\gamma L_s W$.

The regularization enforces spatial smoothness of $W$, while dropout introduces stochasticity. Together, they prevent the soft mode from becoming localized, ensuring that the smallest eigenvector of $A$ remains delocalized—a prerequisite for a diverging correlation length and criticality.  We also study the bahavior of homogeneous network with a mask of circulant structure (Fig. S1).

In addition, if the stimulus connects to only a small subset of neurons, the network has no reason to coordinate globally; thus, the stimulus must be dense, rather than sparse. For simplicity, $g(s)$ was chosen to be uniform across all neurons. Other forms of $g(s)$ were also tested. More details can be found the supplemental method.

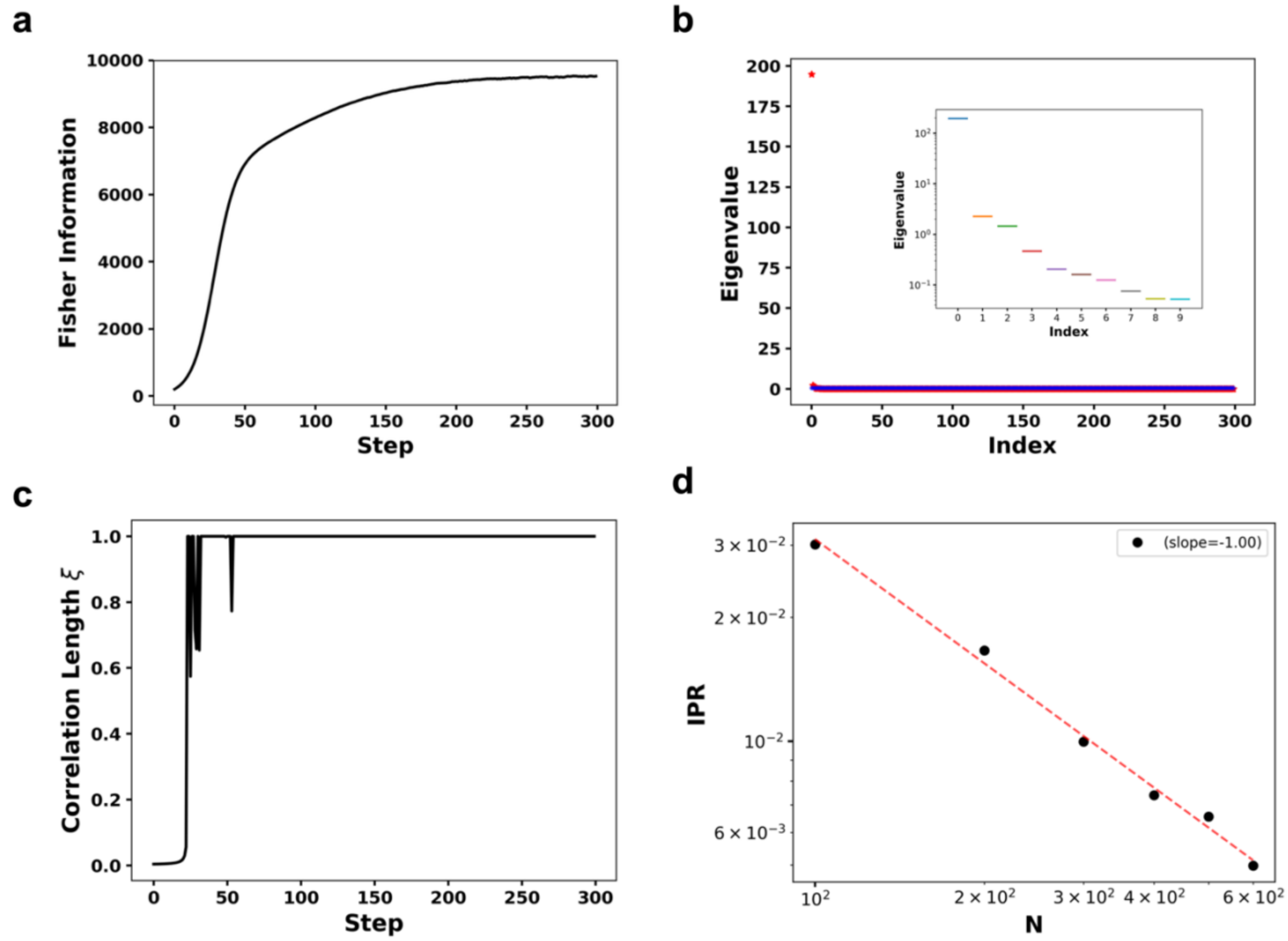

**Figure 3** (a)Evolution of Fisher information over learning (b) distribution of eigenvalue(inserted figure is the largest ten eigenvalues in log scale) (c) Evolution of correlation length over learning (d) IPR with respect to system size.

Initially,Fisher informattion $J(s)$ is low and distributed across many modes. As learning proceeds, Fisher information increases sharply and saturates after approximately 200 iterations (Fig.3a). The increase is accompanied by redistribution of eigenvalues: the largest eigenvalue grows while others shrink. Before learning, eigenvalues are broadly distributed. After learning, the spectrum of $A$ shows one or several dominant eigenvalues, with the remaining eigenvalues compressed near zero (Fig. 3b). Thus, the system becomes stiff along stimulus-sensitive directions and sloppy elsewhere. Comparing with a homogeneous network (Fig. S1), the heterogeneous network exhibits more uneven eigenvalues, indicating that it can achieve higher Fisher information under the same resource constraint, consistent with the learned Fisher information values (Fig. S1).

Since $\lambda_i = 1/\sigma_i^2$, small eigenvalues correspond to large fluctuations along those sloppy eigenvectors. As learning progresses, $\lambda_{\min} \to 0$ due to the resource constraint. The correlation length $\xi$ then diverges, saturating at the system size—a hallmark of statistical criticality (Fig.3c).

To link these sloppy modes to global soft modes [38], we impose spatial structure and constraint in learning. The IPR of softest mode remains delocalized and scales with system size i.e., IPR $\sim N^{-1}$, confirming its global nature (Fig. 3d). In this regime, the neural population no longer behaves as a collection of independent units but as a coherent, long-range correlated system, capable of the large-scale communication. For the homogeneous network, the translation-invariant connectivity also gives rise to a delocalized softest mode (Fig. S2). These conclusions are robust to different forms of global stimuli (Fig. S3).

When the smallest eigenvalue of $A$ approaches zero ($\lambda_{\min} \to 0$), the corresponding eigenvalue of transfer matrix $T$ approaches 1 from below ($\mu_{\max} \to 1^{-}$). This indicates critical slowing down: perturbations along the soft mode decay increasingly slowly, and the system becomes marginally stable. And it is the precursor to a bifurcation [39]. Thus, in this simple Gaussian population coding model, diverging correlation length (statistical criticality) and bifurcation dynamics (dynamical criticality) are two manifestations of the same underlying phenomenon: sloppy efficient coding in nerual networks driven by the optimization of Fisher information under resource constraints.

We interpret the elements of $A$ not as physical or synaptic connections, but as induced effective couplings. Such dense effective interactions naturally arise in systems driven by common latent inputs. While the sturctural or anatomical connectivity $M$ is sparse and local, the resulting effective interacition $A$ can become dense with long-range correlations due to the diverging correlation length of the soft modes. And the precision matrix $A$ also encodes the effective interaction landscape. Soft modes with $\lambda_{soft} \to 0$ create flat directions in the effective energy landscape, thereby facilitating generalization and supporting degenerate solutions near minima [40,41]. The precision matrix $A$ is also related with the Gaussian graphical model. If an entry in the precision matrix is zero, the corresponding variables are conditionally independent given all other variables [42,43].

Beyond the critical slowing down of relaxation dynamics, we also examine system dynamics through its response to a quench event, defined as a rapid fluctuation in the precision matrix. We define the response magnitude, denoted as $\boldsymbol{dx}$, as the shift in the neural population state vector induced by the quench. Such quench events may arise from short-term plasticity and are expected to occur predominantly along sloppy directions [44]. Mathematically, if the system operates at a fixed point

$$x^* = A^{-1}\mathbf{F}$$

determined by the external drive $\mathbf{F}$, a fluctuation $A \rightarrow A + \delta A$ forces the system to migrate to a new equilibrium. To maintain a stable population code where the mean activity corresponds to a biologically realistic tuning curve $f(s)$, the external drive must satisfy the equilibrium condition $\boldsymbol{F}(s) = Af(s)$. Critically, the external drive vector $\mathbf{F}$ is modeled as high-dimensional Gaussian white noise, representing the background synaptic input or spontaneous activity inherent to cortical networks.

The response vector is calculated as the difference between steady states:

$$\mathbf{dx} = (A + \delta A)^{-1}\mathbf{F} - A^{-1}\mathbf{F}$$

In the limit of small fluctuations, this can be approximated via the resolvent expansion as $\boldsymbol{dx} \approx -A^{-1}(\delta A)\boldsymbol{x}^*$. This formulation highlights that the response is dominated by the inverse of the precision matrix $A^{-1}$.

In sloppy systems, the eigenvalue spectrum spans many orders of magnitude, with a dense concentration of near-zero eigenvalues (the soft modes). Because the magnitude of the response scales as $\|\boldsymbol{dx}\| \propto 1/\lambda_{soft}$ projections of random structural noise onto these soft modes, even if the fluctuation $\delta A$ is Gaussian white noise, the resulting distribution of response magnitudes $P(\|\boldsymbol{dx}\|)$ exhibits a heavy-tailed power law (Fig. 4). This power law arises directly from the singular geometry of the precision matrix, effectively translating the spectral density of the effective coupling into a macroscopic statistical distribution. Biologically, this static displacement $\mathbf{dx}$ can serve as a proxy for the scale of neuronal avalanches. Thus, the heavy-tailed distribution of static shifts can be associated with the power-law statistics of avalanche observed in cortex [45], suggesting that neural criticality can arise from the inherent sloppiness of the population coding geometry. In contrast, the response magnitudes of a shuffled precision matrix or a matrix with localized soft modes do not follow a power-law distribution (Fig. S4).

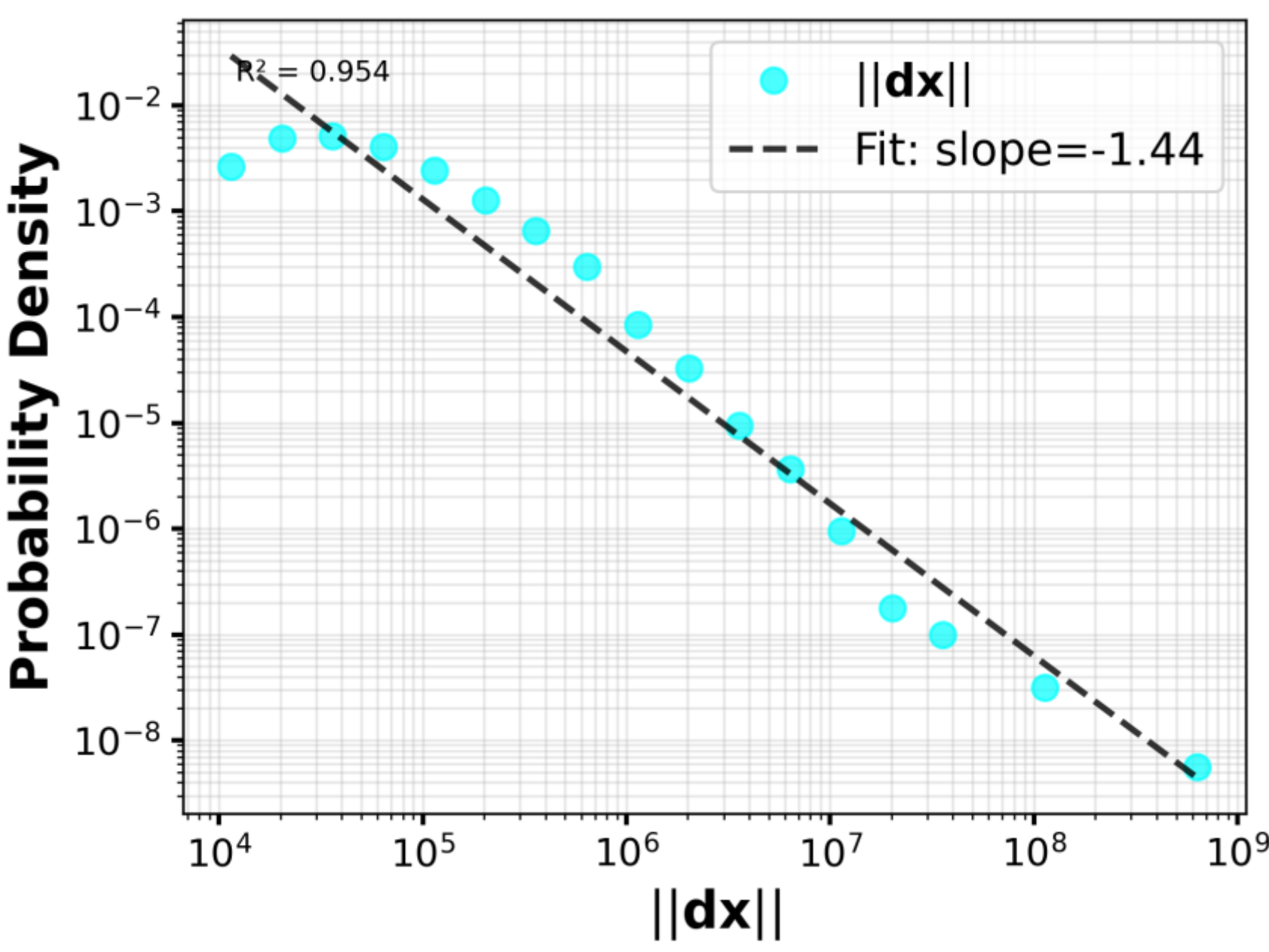

**Figure 4** Distribution of response magnitude in the quech events of precision matrix.

The transition to criticality in our model is not a fine-tuned accident but a consequence of information-theoretic compromise. A non-critical system with uniform eigenvalues would be isotropic but inefficient, wasting energy on non-informative noise. It should be noted that the entropy constraint and network topology settles the system into a quasi-critical state. And this sloppy structure that the spectrum of $A$ is stiff in a few informative dimensions but extremely soft in most others allows the brain to remain stable against random noise while retaining the flexibility to reorganize its global state rapidly in response to relevant stimuli or signals.

## IV. DISCUSSION

While numerous mechanisms—such as frustration from competing interactions[46], excitation-inhibition balance [47], and anti-Hebbian plasticity [48]—have been proposed to explain *how* criticality is generated, our framework explored the complementary question of *why* neural systems are poised at criticality.

One study has suggested a possible relationship between efficient coding and criticality, but the underlying mechanism remains unclear[49]. By integrating Fisher information and sloppiness, [13,14,44,50-52], we have shown that efficient coding under resource constraints drives neural population models towards both criticality and sloppiness. This framework unifies statistical and dynamical perspectives on brain criticality. Our work suggests that brain criticality may be an inevitable consequence of trade-off between precision and energy cost. The trace constraint on $A$ captures metabolic costs of maintaining low neural variability, which may reflecting energy expenditure on ion pumps and neurotransmitter recycling. Robustness is often defined as the ability to maintain function despite internal fluctuations or environmental perturbations. With this model, we show robustness is the necessary flip-side of selective sensitivity. This sloppiness provides biological robustness: the encoded information is immune to noise or synaptic drifts occurring in the sloppy directions [52].

Nevertheless, our model relies on several simplifying assumptions. First, the assumption that the covariance is independent of the stimulus, while analytically tractable, is biologically unrealistic. Future work should consider stimulus-dependent precision to better capture the adaptive nature of neural coding. Second, the stimulus considered here is one-dimensional, whereas real neural systems typically process multidimensional stimuli simultaneously. In that case, scalar Fisher information becomes a Fisher information matrix, and the learning objective should be to maximize its determinant [10]. In addition, the variation of stimuli as well as signal correlation should also be taken into consideration in future work. Finally, this work does not directly explore neural dynamics in avalanche, as we begin from the distribution of neural responses. Bridging the gap between real dynamics governed by synaptic weights and the effective interactions represented by the precision matrix will be critical for understanding the underlying mechanisms of sloppiness and efficient coding.

**ACKNOWLEDGMENT**

This work was supported by National Natural Science Foundation of China Grants No.12247104 to W.W .

**Reference**

[1] P. Bak, C. Tang, and K. Wiesenfeld, Physical Review Letters **59**, 381 (1987).
[2] K. B. Hengen and W. L. Shew, Neuron **113**, 2582 (2025).
[3] J. M. Beggs, Frontiers in Computational Neuroscience **Volume 16 - 2022** (2022).
[4] T. Mora and W. Bialek, Journal of Statistical Physics **144**, 268 (2011).
[5] O. Kinouchi and M. Copelli, Nature Physics **2**, 348 (2006).
[6] L. Barnett, J. T. Lizier, M. Harré, A. K. Seth, and T. Bossomaier, Physical Review Letters **111**, 177203 (2013).
[7] D. Marinazzo, M. Pellicoro, G. Wu, L. Angelini, J. M. Cortés, and S. Stramaglia, PLOS ONE **9**, e93616 (2014).
[8] Z. Wang, A. A. Stocker, and D. D. Lee, in *Neural Information Processing Systems 2013*.
[9] X. X. Wei and A. A. Stocker, Neural Comput **28**, 305 (2016).
[10] N. Brunel and J. P. Nadal, Neural Comput **10**, 1731 (1998).
[11] S. A. FRANK, Journal of Evolutionary Biology **22**, 231 (2009).
[12] J. Hidalgo, J. Grilli, S. Suweis, M. A. Muñoz, J. R. Banavar, and A. Maritan, Proceedings of the National Academy of Sciences **111**, 10095 (2014).
[13] A. Ponce-Alvarez, G. Mochol, A. Hermoso-Mendizabal, J. de la Rocha, and G. Deco, eLife **9**, e53268 (2020).
[14] D. Panas, H. Amin, A. Maccione, O. Muthmann, M. van Rossum, L. Berdondini, and M. H. Hennig, J Neurosci **35**, 8480 (2015).
[15] M. Prokopenko, J. T. Lizier, O. Obst, and X. R. Wang, Physical Review E **84**, 041116 (2011).
[16] W. Janke, D. A. Johnston, and R. Kenna, Physica A: Statistical Mechanics and its Applications **336**, 181 (2004).
[17] B. B. Averbeck, P. E. Latham, and A. Pouget, Nature Reviews Neuroscience **7**, 358 (2006).
[18] J.-B. Eppler, T. Lai, D. F. Aschauer, S. Rumpel, and M. Kaschube, Proceedings of the National Academy of Sciences **123**, e2503046123 (2026).
[19] R. Rafeh and G. Gupta, The Journal of Neuroscience **40**, 7782 (2020).
[20] V. Pernice and R. A. da Silveira, PLOS Computational Biology **14**, e1005979 (2018).
[21] V. R. Bejjanki, J. M. Beck, Z.-L. Lu, and A. Pouget, Nature Neuroscience **14**, 642 (2011).
[22] Y. Gu, S. Liu, Christopher R. Fetsch, Y. Yang, S. Fok, A. Sunkara, Gregory C. DeAngelis, and Dora E. Angelaki, Neuron **71**, 750 (2011).
[23] L. F. Abbott and P. Dayan, Neural Comput **11**, 91 (1999).
[24] L. Aitchison, G. Hennequin, and M. Lengyel, arXiv: Neurons and Cognition (2018).
[25] M. R. Nassar, D. Scott, and A. Bhandari, The Journal of Neuroscience **41**, 6740 (2021).
[26] D. N. Scott and M. J. Frank, bioRxiv, 2021.11.19.466943 (2021).
[27] B. v. Zwol, R. Jefferson, and E. L. v. d. Broek, ACM Comput. Surv. (2026).
[28] M. W. Spratling, Brain and Cognition **112**, 92 (2017).
[29] Z.-Y. Huang, R. Zhou, M. Huang, and H.-J. Zhou, Science China Physics, Mechanics & Astronomy **67**, 260511 (2024).
[30] M. B. Hastings and T. Koma, Communications in Mathematical Physics **265**, 781 (2006).
[31] R. Pastor-Satorras and C. Castellano, Scientific Reports **6**, 18847 (2016).
[32] A. V. Goltsev, S. N. Dorogovtsev, J. G. Oliveira, and J. F. F. Mendes, Physical Review Letters **109**, 128702 (2012).
[33] F. L. Metz, I. Neri, and D. Bollé, Physical Review E **82**, 031135 (2010).

[34] Y. R. Yue and X. Wang, Journal of biometrics & biostatistics **5**, 1 (2013).
[35] F. Lindgren, H. Rue, and J. Lindström, Journal of the Royal Statistical Society Series B: Statistical Methodology **73**, 423 (2011).
[36] D. S. Bassett and E. T. Bullmore, Neuroscientist **23**, 499 (2017).
[37] S. F. Muldoon, E. W. Bridgeford, and D. S. Bassett, Scientific Reports **6**, 22057 (2016).
[38] T. Schneider, G. Srinivasan, and C. P. Enz, Physical Review A **5**, 1528 (1972).
[39] M. G. Crandall and P. H. Rabinowitz, Journal of Functional Analysis **8**, 321 (1971).
[40] Y. Feng and Y. Tu, Proceedings of the National Academy of Sciences **118**, e2015617118 (2021).
[41] C. Baldassi, F. Pittorino, and R. Zecchina, Proceedings of the National Academy of Sciences **117**, 161 (2020).
[42] K. H. Shutta, R. De Vito, D. M. Scholtens, and R. Balasubramanian, Stat Med **41**, 5150 (2022).
[43] Z. Song, S. Gunn, S. Monti, G. M. Peloso, C.-T. Liu, K. Lunetta, and P. Sebastiani, Frontiers in Systems Biology **Volume 5 - 2025** (2025).
[44] M. H. Hennig, The Journal of Physiology **601**, 3025 (2023).
[45] J. M. Beggs and D. Plenz, The Journal of Neuroscience **23**, 11167 (2003).
[46] Y. I. Wolf, M. I. Katsnelson, and E. V. Koonin, Proceedings of the National Academy of Sciences **115**, E8678 (2018).
[47] J. Beggs and N. Timme, Frontiers in Physiology **3** (2012).
[48] M. O. Magnasco, O. Piro, and G. A. Cecchi, Physical Review Letters **102**, 258102 (2009).
[49] S. Safavi, M. Chalk, N. K. Logothetis, and A. Levina, Proceedings of the National Academy of Sciences **121**, e2302730121 (2024).
[50] K. N. Quinn, M. C. Abbott, M. K. Transtrum, B. B. Machta, and J. P. Sethna, Reports on Progress in Physics **86**, 035901 (2023).
[51] M. K. Transtrum, B. B. Machta, K. S. Brown, B. C. Daniels, C. R. Myers, and J. P. Sethna, The Journal of Chemical Physics **143** (2015).
[52] B. C. Daniels, Y.-J. Chen, J. P. Sethna, R. N. Gutenkunst, and C. R. Myers, Current Opinion in Biotechnology **19**, 389 (2008).